\documentclass[runningheads,envcountsame,10pt]{llncs}
\usepackage{times}
\usepackage{eucal}
\usepackage{latexsym}
\usepackage{amsmath}
\usepackage{amssymb}
\newcommand{\person}[1]{\textsc{#1}}

\newcommand{\IR}{\mathbb{R}}
\newcommand{\IN}{\mathbb{N}}

\newcommand{\IC}{\mathbb{C}}
\newcommand{\IH}{\mathbb{H}}
\newcommand{\IK}{\mathbb{K}}
\newcommand{\id}{\operatorname{id}}

\newcommand{\calR}{\mathcal{R}}

\newcommand{\polylog}{\operatorname{polylog}}
\renewcommand{\Re}{\operatorname{Re}}
\renewcommand{\Im}{\operatorname{Im}}
\newcommand{\calO}{\mathcal{O}}
\newcommand{\COMMENTED}[1]{}

\begin{document}
\setcounter{page}{705}
\title{Quasi-Optimal Arithmetic for Quaternion Polynomials}
\author{Martin Ziegler\thanks{Supported 
by \textsf{PaSCo}, DFG Graduate College no.693}}
\authorrunning{M. Ziegler}
\institute{University of Paderborn, 33095 GERMANY; ~\email{ziegler@upb.de}}
\maketitle
\begin{abstract}
Fast algorithms for arithmetic on real or complex
polynomials are well-known and have
proven to be not only asymptotically efficient but also very practical.
Based on \emph{Fast Fourier Transform}, they for instance
multiply two polynomials of degree up to $n$ or multi-evaluate
one at $n$ points simultaneously within quasi-linear time
~$\calO(n\cdot\polylog n)$. An extension to (and in fact
the mere definition of) polynomials over fields $\IR$ and $\IC$
to the \emph{skew}-field $\IH$
of quaternions is promising but still missing.
The present work proposes three approaches which
in the commutative case coincide but for $\IH$ turn out
to differ, each one satisfying some desirable properties while
lacking others. For each notion, we devise algorithms
for according arithmetic; these are quasi-optimal in
that their running times match lower complexity bounds
up to polylogarithmic factors.
\end{abstract}

\section{Motivation}
Nearly 40 years after \person{Cooley} and \person{Tukey} \cite{FFT},
their \textsf{Fast Fourier Transform} (FFT) has provided
numerous applications, among them
\begin{itemize}
\vspace*{-1ex}
\itemsep5pt
\item fast \textsf{multiplication} of polynomials
\begin{quote} \it
Given the coefficients of ~$p,q\in\IC[X]$, 
~$n:=\deg(p)+\deg(q)$; \\ determine the coefficients of $p\cdot q$.
\end{quote}
which, based on FFT, can be performed in ~$\calO(n\cdot\log n)$~
\qquad\qquad\qquad and
\item their \textsf{multi-evaluation}
\begin{quote} \it
Given the coefficients of $p\in\IC[X]$, ~$\deg(p)<n$,
and $x_1,\ldots,x_n\in\IC$; \\ determine the values
$p(x_1), \ldots, p(x_n)$.
\end{quote}
allowing algorithmic solution within $\calO(n\cdot\log^2 n)$.
\end{itemize}
\vspace*{-1ex}
Observe in both cases the significant improvement
over naive ~$\calO(n^2)$ approaches.
These two examples illustrate a larger class of operations
called \emph{Fast Polynomial Arithmetic} \cite{BiniPan,Zippel}
with, again, a vast number of applications \cite{MCA}.
For instance, \person{Gerasoulis} employed fast polynomial
arithmetic to drastically accelerate \textsf{$N$-Body Simulations}
in 2D \cite{Gerasoulis}, and \person{Pan}, \person{Reif}, and
\person{Tate} did so in 3D \cite{ReifTate}. Since systems with
up to $N=10^5$ objects arise quite frequently when simulating
biochemical processes, the theoretical benefit of asymptotic
growth $\calO(N\cdot\polylog N)$ over $\calO(N^2)$
pays off in practice as well.

Technically speaking
in order to calculate, for each of the $N$ particles, 
the total force it experiences due to the $N-1$ others,
\person{Gerasoulis} identifies the plane $\IR^2$ with $\IC$;
he thus turns Coulomb's potential into a rational complex
function which, by means of
fast polynomial multiplication and multi-evaluation, can
be handled efficiently. 
\cite{ReifTate,WADS} on the other hand
exploit fast multi-evaluation of polynomials
to \emph{approximate} the total forces in $\IR^3$.
Whether \textsf{3D} forces can be obtained \emph{exactly}
within subquadratic time is still an open question.
One promising approach proceeds by identifying, similarly
to \cite{Gerasoulis}, $\IR^3$ with (a subspace of)
\person{Hamilton}'s four-dimensional algebra of Quaternions
$\IH$ and there applying fast polynomial arithmetic of some
kind or another. In fact the mere notion of a polynomial becomes
ambiguous when passing from fields $\IK=\IR$ and $\IK=\IC$
to the {skew-}field $\IK=\IH$. We consider three common
approaches to define polynomials (Section~\ref{secQuaternions})
and, for each induced kind of quaternion polynomials, present
quasi-optimal algorithms supporting
according arithmetic operations (Section~\ref{secAlgo}).

\section{Quaternions} \label{secQuaternions}
The algebra $\IH$ of quaternions was discovered in 1843 by
\person{W.R. Hamilton} in an attempt to extend multiplication
of 'vectors' from $\IR^2\cong\IC$ to $\IR^3$. In fact,
$\IH$ is a \emph{four}-dimensional real vector space whose canonical
basis $1,i,j,k$ satisfies the non-commutative multiplicative rule
\begin{equation} \label{eqBasis}
 i^2=j^2=k^2=ijk=-1, \quad\qquad ij=-ji=k \qquad  \text{+ cyclic interchange}
\end{equation}
which, by means of associative and distribute laws, is extended
to arbitrary quaternions.
$\IH$ is easily verified to form a \emph{skew-field}, that is, any
non-zero element $a$ possesses a unique two-sided multiplicative
inverse $a^{-1}$. In fact it holds $a^{-1}=\bar a/|a|^2$ where
$\bar a:=\Re(a)-i\Im_i(a)-j\Im_j(a)-k\Im_k(a)$
is the analogue of complex conjugation and
$|a|:=\sqrt{a\cdot\bar a}=\sqrt{\bar a\cdot a}\in\IR_+$
the \emph{norm} satisfying $|a\cdot b|=|a|\cdot|b|$.
The \emph{center} of $\IH$ is $\IR$; in other words:
real numbers and only they multiplicatively commute with any quaternion.
For further details, please refer to
the excellent\footnote{wrongly condemned
 in \textsc{Chapter XXI, p.245} of \cite{Rota}\ldots}
\textsc{Chapter~7} of \cite{Numbers}.
\textsc{Theorem~17.32} in \cite{ACT} determines the
(multiplicative algebraic) complexity of quaternion multiplication;
\cite{Blaeser} does so similarly for quaternion inversion and division.
However rather than on single quaternions,
our focus shall lie on asymptotics w.r.t. $n$, the
quaternion polynomials' degree, tending to infinity.

It is well-known that commutativity \emph{has} to be abandoned
in order to turn $\IR^4$ into some sort of a field; in fact,
\person{Frobenius}' Theorem states that $\IH$ is the \emph{only}
associative division algebra beyond $\IR^2\cong\IC$. On the other
hand to the author's best knowledge, all notions of polynomials
either require the ground ring $\calR$ to satisfy commutativity or
--- such as \emph{skew polynomial rings}, see \textsc{p.262},
\textsc{Chapter~16} of \cite{Ore} ---
they lack evaluation homomorphisms. The latter
means that any polynomial $p=p(X)\in\calR[X]$ should naturally
induce a mapping $\hat p:\calR\to\calR$, $x\mapsto\hat p(x)$ such that
for all $a,x\in\calR$:
$$\hat X(x)=x, \quad \hat a(x)=a, \quad
\widehat{p\cdot q}(x)=\hat p(x)\cdot\hat q(x), \quad
\text{and } \quad \widehat{p+q}(x)=\hat p(x)+\hat q(x) \enspace . $$

The distant goal is to find a notion of quaternion polynomials
which naturally generalizes from real or complex ones \emph{and}
supports efficient arithmetic by means of, say, quasi-linear time
algorithms. Our contribution considers three such definitions
for $\IK[X]$ which, in case $\IK$ is an infinite field, are
equivalent to the usual notion. In case $\IK=\IH$ however
they disagree and give rise to different arithmetic operations.
We focus on \textsf{Multiplication} and \textsf{Multi-Evaluation}
and present in Section~\ref{secAlgo},
for each of the three notions, according quasi-optimal algorithms.

\subsection{Polynomials as Ring of Mappings} \label{subsecPolyOne}
The idea pursued in this subsection is that
the following objects should be considered polynomials:
\begin{itemize}
\itemsep0pt
\item the identity mapping ~$X:=\id:\IK\to\IK$, ~$x\mapsto x$,
\item any constant mapping ~$\hat a:\IK\to\IK$, ~$x\mapsto a$
       \quad for $a\in\IK$
\item the sum of two polynomials \qquad and
\item the product of two polynomials.
\end{itemize}
Formally, let the set $\IK^\IK$ of mappings $f:\IK\to\IK$ inherit
the ring structure of $\IK$ by defining pointwise
\qquad $f+g:x\mapsto f(x)+g(x)$, \quad $f\cdot g:x\mapsto f(x)\cdot g(x)$.
Then embed $\IK$ into this ring by identifying
$a\in\IK$ with the constant mapping $\IK\ni x\mapsto a\in\IK$.
\begin{definition} \label{defPolyOne}
$\IK_1[X]$ is the smallest subring of $\IK^\IK$ containing
$X$ and the constant mappings $\IK$.  For instance,
\begin{equation} \label{eqExample}
a_1+X\cdot a_2\cdot X\cdot X\cdot a_3+a_4\cdot X\cdot X\cdot X\cdot a_5
 \in\IK_1[X]\;\;, \;\;\quad a_1,\ldots,a_5\in\IK ~\text{ fixed.}\quad
\end{equation}
\end{definition}
$\IK_1[X]$ is closed not only under addition
and multiplication but also under composition,
i.e., $f+g,f\cdot g,f\circ g\in\IK_1[X]$ for $f,g\in\IK_1[X]$.
Since, in the commutative case, any such polynomial can
be brought to the form
\begin{equation} \label{eqPolyNormal} 
\displaystyle\sum\nolimits_{\ell=0}^{n-1} a_\ell X^\ell,
  \qquad n\in\IN, \quad a_\ell\in\IK \enspace ,
\end{equation}
Definition~\ref{defPolyOne} there obviously coincides with the classical
notion of polynomial rings $\IR[X]$ and $\IC[X]$.
For the skew-field $\IK=\IH$ of quaternions,
the structure of $\IH_1[X]$ is not so clear at first sight:
\begin{itemize}
\itemsep0pt
\item $a \cdot\!X\not=X\!\cdot a$ unless $a\in\IR$~
  i.e., the form (\ref{eqPolyNormal}) in general cannot be attained any more.
\item Uniqueness becomes an issue, since
  \begin{equation} \label{eqPolyZero}
  X\cdot X\cdot i\cdot X\cdot i \;+\; i\cdot X\cdot X\cdot i\cdot X
     \;-\; i\cdot X\cdot i\cdot X\cdot X \;-\; X\cdot i\cdot X\cdot X\cdot i
  \quad \end{equation}
   vanishes identically \cite[\textsc{top of p.201}]{Numbers};
	   \\ in particular, a polynomial can have many more roots
   than its 'degree' suggests.
\item The fundamental theorem of algebra is violated as well:
  $i\cdot X-X\cdot i+1$ has no root in $\IH$
  \cite[\textsc{p.205}]{Numbers}.
\item Lagrange-style polynomials $P_m$ to pairwise distinct
  points $x_0,\ldots,x_{n-1}\in\IH$, e.g.,
  $$ \bigg(\prod_{\substack{\ell=0 \\ \ell\not=m}}^{n-1}
  (X-x_\ell)\bigg) \cdot \bigg(\prod_{\substack{\ell=0 \\ \ell\not=m}}^{n-1}
  (x_m-x_\ell)\bigg)^{-1}
    \qquad\text{or}\qquad
     \prod_{\substack{\ell=0 \\ \ell\not=m}}^{n-1}
       \Big( (x_m-x_\ell)^{-1}\cdot(X-x_\ell)\Big) $$
  both interpolate 
  ~$P_m(x_m)=1$, $P_m(x_\ell)=0$, $m\not=\ell$~
  but obviously lack uniqueness.
\item There is no polynomial division with remainder;
  e.g. \hfill $\displaystyle X\!\cdot\!i\!\cdot\!X\!\mod X^2 \;=\;\; ???$
\end{itemize}
\pagebreak

On the other hand we present in Subsection~\ref{subsecAlgoOne}
algorithms for addition, multiplication, and multi-evaluation
of this kind of quaternion polynomials of degree $n$
in time $\calO(n^4\cdot\polylog n)$.
Since it turns out that generic $p\in\IH_1[X]$ have
roughly $n^4$ free coefficients, the running time is thus
quasi-optimal.
Finally, a fast randomized zero-tester for expressions like
(\ref{eqExample}) and (\ref{eqPolyZero}) comes out easily.

\subsection{Polynomials as Sequence of Coefficients} \label{subsecPolyTwo}
Since the above Definition~\ref{defPolyOne} thus does
not \emph{allow} for quaternion polynomial arithmetic as fast
as quasi-linear time,
the present subsection proposes another approach. The idea is to
identify polynomials with their coefficients.
Recall that for
$p=\sum_{\ell=0}^{n-1} a_\ell X^\ell$ and
$q=\sum_{\ell=0}^{m-1} b_\ell X^\ell$
over a commutative field $\IK$, the finite sequence of coefficients
~$\vec c=(c_\ell)\in\IK^*$~
of $p\cdot q$ is given in terms of
~$\vec a=(a_\ell)\in\IK^*$~ and ~$\vec b=(b_\ell)\in\IK^*$~
by the convolution product
\begin{equation} \label{eqConvolution} 
\textstyle
\vec c \;=\; \vec a * \vec b, \quad
c_\ell = \sum\nolimits_{t=0}^\ell  a_t \cdot b_{\ell-t}, \quad
\ell=0,...,n\!+\!m\!-\!1
\end{equation}
with the implicit agreement that $a_\ell=0$ for $\ell\geq n$
and $b_\ell=0$ for $\ell\geq m$.

\begin{definition} \label{defPolyTwo}
$\IK_2[X]$ is the set $\IK^*$ of finite sequences of quaternions,
equipped with componentwise addition and convolution product
according to (\ref{eqConvolution}).
Let $X$ denote the special sequence $(0,1,0,\ldots,0)\in\IK^*$.
\end{definition}
It is easy to see that this turns $\IK_2[X]$ into a ring
which, in case of fields $\IK$ of characteristic zero,
again coincides with the usual ring of polynomials
$\IK[X]$. Here the classical results assert that arithmetic
operations \verb!+! and \verb!*! can be performed within
time $\calO(n)$ and $\calO(n\cdot\log n)$, respectively.
In Subsection~\ref{subsecAlgoTwo}, we show that the
same is possible in the non-commutative ring $\IH_2[X]$.
Dealing with $n$ coefficients, this is trivially quasi-optimal.

Unfortunately fast arithmetic for $\IH_2[X]$ does not include
multi-evaluation, simply because evaluation
(substituting $X$ for some $x\in\IH$) makes no sense
here: One might be tempted to identify $\vec a\in\IH^*$
with the formal expression $\sum_\ell a_\ell X^\ell$
and $\vec b$ with $\sum_\ell b_\ell X^\ell$, but
then $\vec c:=\vec a*\vec b$ does not agree with
$$\textstyle
\Big(\sum a_\ell X^\ell\Big)\cdot\Big(\sum b_\ell X^\ell\Big)
\quad=\quad
  \sum_\ell \sum_{t=0}^\ell a_t \cdot
   \underbrace{X^t \cdot b_{\ell-t}}_{\not= b_{\ell-t} \cdot X^t}
    \cdot X^{\ell-t}
  \quad\not=\quad \sum_\ell c_\ell X^\ell 
\vspace*{-2.4ex}
$$ because of non-commutativity.

The next subsection considers expressions
of the form $\sum a_\ell X^\ell$ as further notion 
of quaternion polynomials. These lack closure under multiplication;
on the other hand, there, multi-evaluation does make sense
and turns out to have classical complexity $\calO(n\cdot\log^2 n)$.

\subsection{One-sided Polynomials} \label{subsecPolyThree}
Roughly speaking, one aims at a subclass of $\IH_1[X]$
where polynomials have only $\calO(n)$ rather than
$\Theta(n^4)$ coefficients and thus give a chance
for operations with quasi-linear complexity.
\begin{definition} \label{defPolyThree}
Let $X:\IK\to\IK$ denote the identity mapping and consider
this class of mappings on $\IK$\rule{0pt}{2.5ex}: \qquad
$\textstyle\IK_3[X] \quad:=\quad \big\{ \sum\nolimits_{\ell=0}^n
  a_\ell X^\ell :
   n\in\IN_0, a_\ell\in\IK \big\} \quad\subseteq\quad \IK^\IK $.
\\[0.7ex]
\noindent The \textsf{degree} of $p\in\IK_3[X]$ is
\quad ~$\displaystyle\deg(p)=\max_{a_\ell\not=0} \ell$, \quad $\deg(0):=-1$.
\end{definition}
Again this coincides for fields $\IK$ of characteristic zero
with the usual notions. For the skew-field of quaternions,
the restriction compared to (\ref{eqExample}) applies that
all coefficients $a_\ell$ must be on the \emph{left} of
powers $X^\ell$. Unfortunately, this prevents $\IH_3[X]$
from being closed under multiplication; fortunately,
$\IH_3[X]$ has the following other nice properties:
\\ \noindent \begin{minipage}[t]{0.35\textwidth}
\begin{itemize}
\itemsep0pt
\item being a real vector space;
\item supports interpolation;
\end{itemize}
\end{minipage}\hfill\begin{minipage}[t]{0.60\textwidth}
\begin{itemize}
\itemsep0pt
\item allows fast multi-evaluation; 
\item a fundamental theorem of algebra holds;
\item polynomials satisfy uniqueness. \hfill Formally:
\end{itemize}
\end{minipage}
\begin{lemma}  \label{lemUniqueness}
Consider $p:=\sum_{\ell=0}^{n-1} a_\ell X^\ell$, ~$a_\ell\in\IH$.
\vspace*{-0.5ex}
\begin{enumerate}
\item Suppose $p(x)=0$ for all $x\in\IH$. Then $a_\ell=0$ for all $\ell$.
\item Nevertheless even $p\not=0$ may have an infinite (and in particular
       unbounded in terms of $p$'s degree) number of roots.
\item If $a_\ell\not=0$ for some $\ell\geq1$, then $p$ has
      at least one root.
\end{enumerate}
\end{lemma}
\begin{proof}
\begin{enumerate}
\item Follows from Lemma~\ref{lemVandermonde}b) by choosing
       $n\geq\deg(p)$ and pairwise distinct $x_0,\ldots,x_{n-1}\in\IR$
       since then, no three are automorphically equivalent.
\item All quaternions $x=i\beta+j\gamma+k\delta$ with
        $\beta,\gamma,\delta\in\IR$ and
         $\beta^2+\gamma^2+\delta^2=1$
      are easily verified zeros of $p:=X^2+1$.
\item Cf. \textsc{p.205} in \cite{Numbers} or see, e.g., \cite{Fundamental}.
\end{enumerate}
\end{proof}
\textsf{Interpolation} is the question of existence and uniqueness,
given $x_0,\ldots,x_{n-1}$ and 
\linebreak $y_0,\ldots,y_{n-1}\in\IK$, of a
polynomial $p\in\IK[X]$ with degree at most $n-1$
satisfying $p(x_\ell)=y_\ell$ for all $\ell=0,\ldots,n-1$.
In the commutative case, both is asserted for
pairwise distinct $x_\ell$. Over quaternions, this
condition does not suffice neither for uniqueness
(Lemma~\ref{lemUniqueness}b) nor for existence:
\begin{example}
No $p=aX^2+bX+c\in\IH_3[X]$ satisfies $p(i)=0=p(j)$, $p(k)=1$.
\end{example}
It turns out that here an additional condition has to be
imposed which, in the commutative case, holds trivially
for distinct $x_\ell$, namely being automorphically inequivalent.%
\vspace*{-0.9ex}
\begin{definition}
Call $a,b\in\IH$ \textsf{automorphically equivalent}
iff $a=u\cdot b\cdot u^{-1}$ for some non-zero $u\in\IH$,
that is, iff
\qquad $\Re(a)=\Re(b) \quad\wedge\quad  |\Im(a)|=|\Im(b)|$ \qquad
where $\Im(a):=i\Im_i(a)+j\Im_j(a)+k\Im_k(a)$.
\end{definition}
This obviously \emph{is} an equivalence relation (reflexivity,
symmetry, transitivity). The name comes from the fact that mappings
$x\mapsto u\cdot x\cdot u^{-1}$ are exactly the
$\IR$-algebra automorphisms of $\IH$;
cf. \cite[\textsc{bottom of p.215}]{Numbers}.
The central result of \cite{Vandermonde} now says:
\begin{lemma} \label{lemVandermonde}
For $x_0,\ldots,x_{n-1}\in\IH$, the following are equivalent%
\vspace*{-0.5ex}
\begin{enumerate}
\item To any $y_0,\ldots,y_{n-1}\in\IH$, there exists $p\in\IH_3[X]$
      of $\deg(p)<n$ such that $p(x_\ell)=y_\ell$, $\ell=0,\ldots,n-1$.
\item Whenever $p=\sum_{\ell=0}^{n-1} a_\ell X^\ell$ and
      $q=\sum_{\ell=0}^{n-1} b_\ell X^\ell$ satisfy
      $p(x_\ell)=q(x_\ell)$ for $\ell=0,\ldots,n-1$,
      it follows $a_\ell=b_\ell$.
\item The \textsf{Quaternion Vandermonde Matrix}
       ~$\displaystyle V:=(x_\ell^m)_{\scriptscriptstyle\ell,m=0,..,n-1}$
      is invertible.
\item Its \textsf{Double Determinant} ~ $\|V\|$ does not vanish.
\item The $x_\ell$ are pairwise distinct \textsf{and}
       no three of them are automorphically equivalent.
\end{enumerate}
\end{lemma}
Concluding this subsection, $\IH_3[X]$ has
(unfortunately apart from closure under multiplication)
several nice structural properties.
In \ref{subsecAlgoThree} we will furthermore show that it
supports multi-evaluation in time $\calO(n\cdot\log^2 n)$.
More generally, our algorithm applies to polynomials
\hfill $\textstyle \IH^1_3[X] \quad := \quad
  \big\{ \sum\nolimits_{\ell=0}^{n-1} a_\ell\cdot X^\ell\cdot b_\ell\;:\;
         n\in\IN_0, \; a_\ell,b_\ell\in\IH \big\}$
with coefficients to \emph{both} sides of each monomial $X^\ell$.
This generalized notion has the advantage of yielding not only an
$\IR$-vector space but a two-sided $\IH$-vector space.

\section{Algorithms}  \label{secAlgo}
\subsection{Convolution of Quaternion Sequences} \label{subsecAlgoTwo}
Beginning with the simplest case of $\IH_2[X]$: \\
Let $n\in\IN$. Given $\vec a=(a_0,a_1,\ldots,a_{n-1})\in\IH^{n}$
and $\vec b=(b_0,b_1,\ldots,b_{m-1})\in\IH^{m}$, one can
compute their convolution according to (\ref{eqConvolution})
from 16 real convolutions\footnote{In fact, 4 complex convolutions 
suffice; but asymptotically, that gains nothing.}
and 12 additions of real sequences within time $\calO(n\cdot\log n)$.
Indeed write componentwise
$$ \textstyle
 \vec a \!=\! \Re(\vec a)+ i\Im_i(\vec a)+
   j\Im_j(\vec a)+k\Im_k(\vec a), \;\;
 \vec b \!=\! \Re(\vec b)+ i\Im_i(\vec b)+
   j\Im_j(\vec b)+k\Im_k(\vec b) 
$$
and exploit $\IR$-bilinearity of quaternion convolution.
\subsection{Ring of Quaternion Mappings} \label{subsecAlgoOne}
The central point of this subsection is the identification of
$\IH_1[X]$ with the four-fold Cartesian product
of four-variate real polynomials
~$\prod^4\IR[X_0,X_1,X_2,X_3]$.
Formally consider, for ~$f:\IH\to\IH$,~ 
the quadruple $\tilde f$ of four-variate real functions defined by
\begin{equation}\label{eqComponents}
\hspace*{-1ex}
\begin{array}{r@{\mbox{\scriptsize$:=$}}l@{\,\;\;}r@{\mbox{\scriptsize$:=$}}l}
\mbox{\scriptsize$
\tilde f_0(X_0,..,X_3)
$} &\mbox{\scriptsize$
\Re\big(p(X_0\!+\!iX_1\!+\!jX_2\!+\!kX_3)\big)
$} & \mbox{\scriptsize$
\tilde f_1(X_0,..,X_3) 
$} &\mbox{\scriptsize$
\Im_i\!\!\big(p(X_0\!+\!iX_1\!+\!jX_2\!+\!kX_3)\big)
$} \\ \mbox{\scriptsize$
\tilde f_2(X_0,..,X_3) 
$} &\mbox{\scriptsize$
\Im_j\!\!\big(p(X_0\!+\!iX_1\!+\!jX_2\!+\!kX_3)\big)
$} & \mbox{\scriptsize$
\tilde f_3(X_0,..,X_3)
$} &\mbox{\scriptsize$
\Im_k\!\!\big(p(X_0\!+\!iX_1\!+\!jX_2\!+\!kX_3)\big)
$}
\end{array}\end{equation}
and multiplication among such mappings
$\tilde f,\tilde g:\IR^4\to\IR^4$ given pointwise by
\newcommand{\eqProdOne}{\text{6${\tfrac{1}{2}}$}}
\begin{multline} 
   (\tilde f_0,\tilde f_1,\tilde f_2,\tilde f_3)\,\cdot\,
   (\tilde g_0,\tilde g_1,\tilde g_2,\tilde g_3)
\quad:=\quad  \hspace*{\fill} \makebox{(\eqProdOne)} \\
\scriptstyle
( \tilde f_0\tilde g_0-\tilde f_1\tilde g_1
      -\tilde f_2\tilde g_2-\tilde f_3\tilde g_3,\;
   \tilde f_0\tilde g_1+\tilde f_1\tilde g_0
      +\tilde f_2\tilde g_3-\tilde f_3\tilde g_2,\;
   \tilde f_0\tilde g_2+\tilde f_2\tilde g_0
      +\tilde f_3\tilde g_1-\tilde f_1\tilde g_3,\;
   \tilde f_0\tilde g_3+\tilde f_3\tilde g_0
      +\tilde f_1\tilde g_2-\tilde f_2\tilde g_1)  \nonumber
\end{multline}
In that way, calculations in $\IH_1[X]$ can obviously
be as well performed in $\prod\limits^4\IR[X_0,..,X_3]$.
This allows for application of classical algorithms for
multivariate polynomials over commutative fields.
But before, we need a notion of \emph{degree} on $\IH_1[X]$:
\begin{definition}  \label{defDegree}
For a commutative multi-variate polynomial, let
$\deg$ denotes its \emph{total} degree; 
e.g., $\deg(x^2y^3)=5$, $\deg(0)=-\infty$.
The \textsf{degree} $\deg(q)$ of a 
quaternion polynomial $q\in\IH_1[X]$ is half the
total degree of the real four-variate polynomial
~$\tilde f_0^2+\ldots+\tilde f_3^2$~
with $\tilde f_0,\ldots,\tilde f_3$ according to (\ref{eqComponents}).
\end{definition}
Rather than the total degree, one
might as well have considered the maximum one
$\deg(x^2y^3):=3$ since, for 4 variables, 
they differ by at most a constant factor.
However we shall later exploit the equality
~$\deg(p\cdot q)\pmb{=}\deg(p)+\deg(q)$~
valid for the first
whereas for the latter in general only the \emph{in}equality
~$\deg(p\cdot q)\pmb{\leq}\deg(p)+\deg(q)$~
holds.
In fact, this nice property carries over to the degree of
quaternion polynomials:
\begin{lemma} \label{lemDegree}
The degree ~$\deg(p)$~ of ~$p\in\IH_1[X]$~ is always integral.
Furthermore it holds ~$\deg(p\cdot q)=\deg(p)+\deg(q)$.
\end{lemma}
Now recall the following classical results on four-variate polynomials:
\begin{lemma} \label{lemClassical}
\begin{enumerate}
\item Given (the coefficients of) $p,q\in\IC[X_0,\ldots,X_3]$,
     the (coefficients of the) product $p\cdot q$
     can be computed in time $\calO(n^4\cdot\log n)$ where
       $n:=\deg(p\cdot q)=\deg(p)+\deg(q)$.
\item Given $p\in\IC[X_0,\ldots,X_3]$ of degree $n$,
  one can compute within $\calO(n^4\cdot\log n)$ steps
  the coefficients of
  ~$p\big(T\cdot(X_0,...,X_3)^\dag+\vec y\big)\in\IC[X_0,...,X_3]$,~
  that is, perform on $p$ an affine variable substitution
  given by $T\in\IC^{4\times 4}$ and $\vec y\in\IC^4$.
\item A given polynomial $p\in\IC[X_0,\ldots,X_3]$
  of degree $n:=\deg(p)$
  can be evaluated on all $n^4$ points of a 4-dimensional complex grid
  $G:=A_0\times A_1\times A_2\times A_3$ such that $A_\ell\subseteq\IC$,
  $|A_\ell|=n$, within time $\calO(n^4\cdot\log^2 n)$.
\item The same holds for the regular affine image
  $G'=T\cdot G+\vec y$~ of such a grid, i.e.,
  $$  G' \;\;=\;\;
     \big\{ T\cdot\vec x+\vec y: \vec x=(x_0,\ldots,x_3)^\dag\in G\big\},
    \qquad T\in\IC^{4\times 4} \text{ regular},
      \quad \vec y\in\IC^4 \enspace . $$
\item Let $p\in\IC[X_0,\ldots,X_3]$ be non-zero, $n\geq\deg(p)$.
  Fix arbitrary $A\subseteq\IC$ of size $|A|\geq 2n$.
  Then, for $(x_0,\ldots,x_3)\in A^4$ chosen uniformly at random,
  the probability of $p(x_0,\ldots,x_3)=0$ is strictly less than
  $\tfrac{1}{2}$.
\end{enumerate}
\end{lemma}
\begin{proof}
\begin{enumerate}
\item Reduction to the univariate case by means of \person{Kronecker}'s
    embedding: cf.~\textsc{Equation~(8.3)} on \textsc{p.62} of \cite{BiniPan}
    for $m:=4$; 
   dealing with the complex
    field $\IC$ rather than an arbitrary ring $\calR$ of coefficients,
    the $\log\!\log$-factor may be omitted.
\item Folklore. A proof had to be removed from the final version
  due to space limitations.
\item Cf.~\textsc{Equation~(8.5)} and the one below
  on \textsc{p.63} of \cite{BiniPan} for $m:=4$, $c:=n$.
\item follows from b).
  It is not known whether multi-evaluation is feasible
  on \emph{arbitrarily} placed $n^4$ points within
  time $\calO(n^4\cdot\polylog n)$.
\item Cf.~\textsc{Subsection~12.1} in \cite{Zippel}.
\end{enumerate}
\end{proof}
One could of course identify in a similar way complex
univariate polynomials $p\in\IC[Z]$
with tuples $p_0,p_1\in\IR[X,Y]$ of real bivariate polynomials.
However the thus obtained
running times of $\calO(n^2\cdot\polylog n)$ thus obtained
for $\IC[Z]$ are strikingly suboptimal, basically because
\emph{not every} tuple of real bivariate polynomials
corresponds to a complex univariate polynomial. For instance,
$z\mapsto\Re(z)$ is well-known not only to
be no complex polynomial but to even violate
\person{Riemann}-\person{Jacoby}'s
equations of complex differentiability.
Surprisingly for quaternion polynomials,
the situation is very different:
\smallskip
\begin{lemma} \label{lemSurjective}
$\Re(X)=\tfrac{1}{4}(X-iXi-jXj-kXk)\in\IH_1[X]$.
More generally,
\textsf{every} quadruple of real four-variate polynomials
corresponds to a quaternion polynomial.
\end{lemma}
The generic quaternion polynomial of degree $n$ thus
has $\Theta(n^4)$ free coefficients. 
Lemmas~\ref{lemClassical} and \ref{lemSurjective} together yield
\begin{theorem} \label{thAlgoOne}
\begin{enumerate}
\item
Multiplication of two quaternion polynomials $p,q\in\IH_1[X]$
is possible in time $\calO(n^4\cdot\log n)$ where
$n:=\deg(p\cdot q)=\deg(p)+\deg(q)$.
\item
Multi-evaluation of $p$ at $x_0,\ldots,x_{n-1}\in\IH$
can be done within $\calO(n^4\cdot\log^2 n)$, $n:=\deg(p)$.
\item
Within the same time, multi-evaluation is even feasible at
as many as $n^4$ points $x$, provided they lie on
a (possibly affinely transformed) ~ $n^4$-grid ~$G$.
\end{enumerate}
\noindent
The above complexities are optimal up to the (poly-)logarithmic factor.
\end{theorem}
Theorem~\ref{thAlgoOne} presumes the polynomial(s) to be given as
(coefficients of four) real four-variate polynomials. But how fast
can one convert input in more practical format like (\ref{eqExample})
or (\ref{eqPolyZero}) to that form? By means of fast multiplication
of \emph{several} polynomials, this can be done efficiently as well:
\begin{theorem}  \label{thExpression}
\begin{enumerate}
\item The (ordered!) product $\prod_{\ell=1}^m p_\ell$
    of $m$ quaternion polynomials $p_\ell\in\IH_1[X]$,
    each given as quadruple of real four-variate polynomials,
    can be computed within $\calO(n^4\cdot\log n\cdot\log m)$
    where $n=\sum_{\ell} \deg(p_\ell)$ denotes the result's degree.
\item An algebraic expression $E$ over quaternions, i.e.,
    composed from $+$, $-$, $\cdot\,$, constants $a\in\IH$,
    and the quaternion variable $X$ 
    --- but \emph{without} powers like $X^{99}$ nor brackets! ---
    can be converted into the quadruple of real four-variate polynomials
    according to (\ref{eqComponents}) within time
    $\calO(N^4\cdot\log^2 N)$ where $N=|E|$ denotes the 
    input string's length.
\end{enumerate}
\end{theorem}
The above conversion yields a deterministic $\calO(N^4\cdot\log^2 N)$-test
for deciding whether a given quaternion expression like (\ref{eqPolyZero})
represents the zero polynomial. When satisfied with a \emph{randomized}
test, the same can be achieved much faster:
\\[0.5ex] {\bf Theorem~\ref{thExpression} (continued)} {\it
\begin{enumerate}
\vspace*{-1.5ex}
\addtocounter{enumi}{2}
\item Given $\varepsilon>0$ and an expression $E$ of length $N=|E|$,
  composed from "\verb!+!", "\verb!-!", "~$\cdot\,$",
  constants $a\in\IH$, the quaternion variable $X$, \textsf{and}
  possibly brackets "\verb!(!", "\verb!)!"; then one can test
  with one-sided error probability at most $\varepsilon$ whether $E$
  represents the zero-polynomial within time
  $\calO(N\cdot\log\tfrac{1}{\varepsilon})$.
\end{enumerate} }
\begin{proof}
\vspace*{-1ex}
\begin{enumerate}
\item Standard \textsf{divide-and-conquer} w.r.t. $m$
  similar to \textsc{Corollary~2.15} in \cite{ACT}.
\item Lacking brackets, the input string $E$ necessarily has the form
\\
  \centerline{$E \quad=\quad  E_1 \;\pm\; E_2 \;\pm\; \ldots \;\pm\; E_M $}
  where $E_\ell$ describes a product $P_\ell$ of quaternion constants
  (degree 0) and the indeterminate $X$ (degree 1). Since obviously
  $\deg(P_\ell)\leq N_\ell:=|E_\ell|$, its real four-variate
  representation is obtainable within $\calO(N_\ell^4\cdot\log^2 N_\ell)$
  steps. Doing so for all $\ell=1,\ldots,M$ leads to running time
  $\calO(N^4\cdot\log^2 N)$ as $\sum_\ell N_\ell\leq N$.
\\
  W.l.o.g. let $\deg(P_1)\leq\deg(P_2)\leq\ldots\leq\deg(P_M)$.
  Adding up the just obtained four-variate representations
  in this increasing order takes additional time
    \quad $\calO(N_1^4+N_2^4+\ldots+N_M^4) \;\leq\;\calO(N^4)$.
\item
  By virtue of standard amplification it suffices to deal
  with the case $\varepsilon=\tfrac{1}{2}$.
  The algorithm considers any
  set $A\subseteq\IR$ of size $|A|\geq 2N$. It chooses
  $x_0,x_1,x_2,x_3\in A$ uniformly and independently at random;
  and then evaluates the input expression $E$ by substituting
  $X:=x_0+ix_1+jx_2+kx_3$. If the result is zero, the algorithm
  reports \texttt{zero}, otherwise \texttt{non-zero}.
\\
  The running time for evaluation is obviously linear in $|E|=N$.
  Moreover, only one-sided errors occur.
  So suppose $E$ represents non-zero $p\in\IH_1[X]$. Then obviously
  $\deg(p)\leq N$ and at least one of the four real four-variate
  polynomials $\tilde p_0,\ldots,\tilde p_3$ 
  according to (\ref{eqComponents}) is non-zero
  as well. By virtue of Lemma~\ref{lemClassical}e), this will be
  witnessed by $(x_0,\ldots,x_3)$
  --- i.e., $p(x_0+ix_1+jx_2+kx_3)\not=0$ ---
  with probability at least $\tfrac{1}{2}$. \qed
\end{enumerate}
\end{proof}

\subsection{Multi-Evaluating Two-Sided Polynomials}  
\label{subsecAlgoThree}
Consider an expression of the form
$p(X)=\sum_{\ell=0}^{n-1} a_\ell X^\ell b_\ell$, \quad $a_\ell,b_\ell\in\IH$.
Expanding $a_\ell=\Re(a_\ell)+i\Im_i(a_\ell)+j\Im_j(a_\ell)+k\Im_k(a_\ell)$
and similarly for $b_\ell$, one obtains, by virtue of distributive laws
and since whole $\IR$ commutes with $X^\ell$, that
it suffices to multi-evaluate expressions of the form
\begin{equation} \label{eqPolyThree} \textstyle
q(X) \quad = \quad 
  \sum\nolimits_{\ell=0}^{n-1}
    \alpha_\ell X^\ell 
  , \qquad \alpha_\ell\in\IR  ~(!)
\end{equation}
since $p(X)$ can be obtained from 16 of them, each multiplied
both from left and right with some basis element $1,i,j,k$.
Now with real $\alpha_\ell$, multi-evaluation of (\ref{eqPolyThree})
is of course trivial on $x_0,\ldots,x_{n-1}\in\IC$; but we want
$x_\ell$ to be arbitrary quaternions! Fortunately, the latter can
efficiently be reduced to the first.

To this end, consider mappings ~$\varphi_u:\IH\to\IH$,
$x\mapsto u\cdot x\cdot u^{-1}$~ with $u\in\IH$ of norm $|u|=1$.
It is well-known \cite[\textsc{pp.214-216}]{Numbers} that,
identifying $\IH$ with $\IR^4$, $\varphi_u$ describes a rotation,
i.e., $\varphi_u\in\operatorname{SO}(\IR^4)$.
Furthermore, restricted to the set
$$\Im\IH  \quad:=\quad \big\{ x\in\IH : \Re(x)=0 \big\}
  \quad\cong\quad\IR^3 $$
of \emph{purely imaginary} quaternions, $\varphi_u$ exhausts whole
$\operatorname{SO}(\IR^3)$ as $u$ runs through all unit quaternions;
this is called \textsf{\person{Hamilton}'s Theorem}.
Finally, $\varphi_u$ is an (and in fact, again, the most general)
$\IR$-algebra automorphism, i.e., satisfies for
  $\alpha\in\IR$ and $x,y\in\IH$:
$$ \varphi_u(\alpha)=\alpha, \quad
  \varphi_u(x+y)=\varphi_u(x)+\varphi_u(y), \quad
  \varphi_u(x\cdot y)=\varphi_u(x)\cdot\varphi_u(y) \enspace . $$
\begin{lemma} \label{lemReduce}
For $v,w\in\Im\IH$, $|v|=1=|w|$, let $u:=(v+w)/|v+w|$;
then $\varphi_u(v)=w$. In particular for $x\in\IH\setminus\IR$,
$v:=\Im(x)/|\Im(x)|$, $w:=i$, it holds
\quad $q(x)=u^{-1}\cdot q(y)\cdot u$ \quad
where $y:=u\cdot x\cdot u^{-1}\in\IR+i\IR\cong\IC$.
\end{lemma}
Our algorithm evaluates $q\in\IR[X]$ simultaneously at $x_1,..,x_n\in\IH$
as follows:
\begin{itemize}
\item For all $x_\ell\in\IR+i\IR$, let $u_\ell:=1$;
\item for each $x_\ell\not\in\IR$, compute (in constant time) $u_\ell$
    according to Lemma~\ref{lemReduce}.
\item Perform in linear time the transformation ~
      $y_\ell:=u_\ell\cdot x_\ell\cdot u_\ell^{-1}$.
\item Use classical techniques to multi-evaluate $q$ at
  $y_1,\ldots,y_n\in\IC$ within $\calO(n\cdot\log^2 n)$.
\item Re-transform the values $q(y_\ell)$ to
        $q(x_\ell)=u_\ell^{-1}\cdot q(y_\ell)\cdot u_\ell$.
\end{itemize}
This proves the claimed running time of $\calO(n\cdot\log^2 n)$. \qed

\section{Conclusion}  \label{secConclusion}
We proposed three generalizations for the notion '\emph{polynomial}'
from fields $\IR$ and $\IC$ to the skew-field $\IH$ of quaternions
and analyzed their respective properties. For each notion, we then
investigated (where applicable) on the algebraic complexity of
operations \textsf{multiplication} and \textsf{multi-evaluation}
on polynomials in terms of their degree. The upper bounds
attained by our respective algorithms match (usually trivial)
lower bounds up to polylogarithmic factors.

However since each of the above notions lacks one (e.g., closure
under multiplication) or another (e.g., quasi-linear
complexity) desirable property, a satisfactory definition for
quaternion polynomials is still missing. Here comes another one,
generalizing the representation of complex polynomials in terms
of their roots:
\begin{equation} \label{eqRoots}
\IK_4[X] \quad := \quad \big\{ a_0 \cdot
(X-a_1)\cdot(X-a_2)\cdots(X-a_n) : n\in\IN_0, a_\ell\in\IH \big\} 
\end{equation}
So what is the complexity for multi-evaluation in $\IH_4[X]$?

In view of the planar $N$-body problem, \person{Gerasoulis}' major
break-through was fast multi-evaluation of complex rational functions
\begin{equation}  \label{eqNBody} 
  \sum_{\ell=1}^N  (X-a_\ell)^{-1}
\end{equation}
for given $a_1,\ldots,a_N\in\IC$ at given $x_1,\ldots,x_N\in\IC$;
cf. also \textsc{Corollary~7} in \cite{WADS}.
Our techniques from Subsection~\ref{subsecAlgoThree} yield
the same for ~$x_\ell\in\IH$~ and ~$a_\ell\in\IR$.
Thus the crucial question remains
whether (\ref{eqNBody}) also allows multi-evaluation
in sub-quadratic time for both $a_\ell$ and $x_\ell$ being quaternions.
But what is a rational quaternion function, anyway?
We do not even know what a quaternion
polynomial is! Observe that, lacking commutativity,
$$ \frac{1}{X-a} + \frac{1}{X-b}
  \quad = \quad \frac{1}{X-a}\cdot\frac{1}{X-b} \cdot (X-b)
    + (X-a)\cdot\frac{1}{X-a}\cdot\frac{1}{X-b} $$
cannot be collected into one single fraction,
in spite of the common denominator.

\subsubsection*{Acknowledgments:}
The idea to use quaternions for $N$-Body Simulation
was suggested by \person{Peter B\"{u}rgisser}.
The author wishes to thank his student,
\person{Tomas Braj\-kovic}, for having chosen him as supervisor. In fact,
Sections~\ref{subsecPolyOne} and \ref{subsecAlgoOne} constitute the
core of \person{Tomas}'
\emph{Staatsexamensarbeit} (high school teacher's thesis).


\newpage
\begin{appendix}
\section{Postponed Proofs}
Here, we collect some proofs which, in the printed
version, had to be quelled due to page constraints.

\begin{lemma} \label{lemIntegral}
Let $f,g\in\IR[X_1,\ldots,X_d]$ denote $d$-variate real polynomials.
Then,
$$\deg(f^2+g^2)=\max\{\deg(f^2),\deg(g^2)\}=2\max\{\deg f,\deg g\}
\enspace . $$
In particular, the total degree of $f^2+g^2$ is even.
\end{lemma}
\begin{proof}
The second equation is trivial, regarding that the total degree
satisfies $\deg(f^2)=2\deg(f)$; similarly for the \emph{in}equality
$\deg(f^2+g^2)\leq\max\{\deg(f^2),\deg(g^2)\}$.
One thus has to show that,
although in $f^2+g^2$ certain terms of coinciding maximum total degree
might indeed cancel, the above inequality is in fact an equality.
This is where the real ground field comes into play:
choose in $f$ and $g$ respective terms $M$ and $N$ of
coinciding maximum total degree,
that is, $M=a\cdot X_1^{m_1}\cdots X_d^{m_d}$ ~and~
$N=b\cdot X_1^{n_1}\cdots X_d^{n_d}$~ with
$\sum m_\ell=\deg f=\deg g=\sum n_\ell$
and $a,b\not=0$.
Then both $M^2=a^2\cdot X_1^{2m_1}\cdots X_d^{2m_d}$ ~and~
$N^2=b^2\cdot X_1^{2n_1}\cdots X_d^{2n_d}$
~have total degree equal to $\deg(f^2)=\deg(g^2)$.
Furthermore, their respective occurrences in $f^2+g^2$ cannot
cancel because $a^2$ and $b^2$ are strictly positive;
hence $\deg(f^2+g^2)\geq \deg(M^2)=\deg(N^2)$.
\qed \end{proof}

\noindent {\bf Proof of Lemma~\ref{lemDegree}}.
Integrality of the degree is covered by
applying Lemma~\ref{lemIntegral} above inductively to
$(\tilde f_0^2,\tilde f_1^2)$,
$(\tilde f_0^2+\tilde f_1^2,\tilde f_2^2)$, and
$(\tilde f_0^2+\tilde f_1^2+\tilde f_2^2,\tilde f_3^2)$.
\\
For the second claim, straight-forward calculation confirms the
\emph{Four Squares Theorem}\footnote{discovered
by \person{Euler} in 1748}
for real numbers, here applied to the case of real
polynomials $\tilde f_0,\ldots,\tilde f_3,\tilde g_0,\ldots,\tilde g_3$:
\begin{eqnarray}
\lefteqn{ \big(\tilde f_0^2+\tilde f_1^2+\tilde f_2^2+\tilde f_3^2\big)\cdot
  \big(\tilde g_0^2+\tilde g_1^2+\tilde g_2^2+\tilde g_3^2\big) \quad=\quad } \nonumber \\
&=& (\tilde f_0\tilde g_0 - \tilde f_1\tilde g_1-\tilde f_2\tilde g_2-\tilde f_3\tilde g_3)^2 \;+\;
    (\tilde f_0\tilde g_1+\tilde f_1\tilde g_0+\tilde f_2\tilde g_3-\tilde f_3\tilde g_2)^2 \label{eqEuler} \\
&+& (\tilde f_0\tilde g_2+\tilde f_2\tilde g_0+\tilde f_3\tilde g_1-\tilde f_1\tilde g_3)^2 \;+\;
    (\tilde f_0\tilde g_3+\tilde f_3\tilde g_0+\tilde f_1\tilde g_2-\tilde f_2\tilde g_1)^2  \nonumber
\end{eqnarray}
Now observe that the total degree of the left hand side
of (\ref{eqEuler}) is
\begin{multline*}
\deg\big((\tilde f_0^2+\tilde f_1^2+\tilde f_2^2+\tilde f_3^2)\cdot(
  (\tilde g_0^2+\tilde g_1^2+\tilde g_2^2+\tilde g_3^2)\big)  
\\ =\quad
  \deg(\tilde f_0^2+\tilde f_1^2+\tilde f_2^2+\tilde f_3^2)
  \:+\: \deg(\tilde g_0^2+\tilde g_1^2+\tilde g_2^2+\tilde g_3^2)
\end{multline*}
which, by Definition~\ref{defDegree}, agrees with
~$2\deg(p)+2\deg(q)$~ for $p,q\in\IH_1[X]$ according to (\ref{eqComponents}).
At the same time, in view of Equation (\eqProdOne),
the total degree of (\ref{eqEuler})'s right hand side
is nothing but ~$2\deg(p\cdot q)$.
\qed 

\medskip\medskip
\noindent {\bf Proof of Lemma~\ref{lemSurjective}}.
\begin{itemize}
\renewcommand{\labelitemi}{\ensuremath{\bullet}}
\item
Straight forward calculation verifies
~$\Re(X)=\tfrac{1}{4}(X-iXi-jXj-kXk)$ which
obviously belongs to $\IH_1[X]$.
Thus, the quadruple of four-variate real polynomials
$(X_0,0,0,0)\in\prod^4\IR[X_0,\ldots,X_3]$
does correspond to a quaternion polynomial.
\item
Similarly, $(X_1,0,0,0)\in\prod^4\IR[X_0,\ldots,X_3]$
corresponds to $\Im_i(X)=\Re(-iX)\in\IH_1[X]$;
same for $(X_2,0,0,0)$ and $(X_3,0,0,0)$.
\item
For any real constant $\alpha$,
$(\alpha,0,0,0)\in\prod^4\IR[X_0,\ldots,X_3]$
corresponds to $\alpha\in\IH\subseteq\IH_1[X]$.
\item
  Let $(\tilde f,0,0,0)$ and $(\tilde g,0,0,0)$ correspond
  to quaternion polynomials $p,q\in\IH_1[X]$, respectively.
  Then $(\tilde f+\tilde g,0,0,0)$ corresponds to $p+q\in\IH_1[X]$; and,
  $$ (\tilde f\cdot \tilde g,0,0,0) \quad\overset{(\eqProdOne)}{=}\quad
     (\tilde f,0,0,0)\cdot(\tilde g,0,0,0) $$
  corresponds to $p\cdot q\in\IH_1[X]$.
\end{itemize}
Since $\IR[X_0,\ldots,X_3]$ is the \emph{smallest} set containing real
constants, the generators $X_0$, $X_1$, $\ldots$, $X_3$, and being closed 
under addition and multiplication, the above considerations imply that,
for any $\tilde f\in\IR[X_0,\ldots,X_3]$, $(\tilde f,0,0,0)$ corresponds
to some $p\in\IH_1[X]$.
\begin{itemize}
\renewcommand{\labelitemi}{\ensuremath{\bullet}}
\item
  Suppose $(\tilde f,0,0,0)$ corresponds to $p\in\IH_1[X]$. Then
  $(0,\tilde f,0,0)$ corresponds to $-ip\in\IH_1[X]$;
  analogously for $(0,0,\tilde f,0)$ and $(0,0,0,\tilde f)$.
\item
  Let $(\tilde f_0,0,0,0)$ correspond to $p_0\in\IH_1[X]$,
  $(0,\tilde f_1,0,0)$ to $p_1$,
  $(0,0,\tilde f_2,0)$ to $p_2$, and $(0,0,0,\tilde f_3)$ to $p_3$.
  Then
  $(\tilde f_0,\tilde f_1,\tilde f_2,\tilde f_3)$
  corresponds to $p_0+\ldots+p_3\in\IH_1[X]$.
\qed
\end{itemize}

\medskip\medskip
\noindent {\bf Proof of Lemma~\ref{lemReduce}}.
Observe that
~$u^{-1}=\bar u/|u|^2=-u$~ since $u\in\Im\IH$ and $|u|=1$. Thus
\begin{eqnarray*} \varphi_u(v)
&=& -\frac{(v+w)v(v+w)}{|v+w|^2}
 \quad =\quad -\frac{v^3+v^2w+wv^2+wvw}{2+2\langle v,w\rangle}
\end{eqnarray*}
because $|v|=1=|w|$ by presumption. As $v,w\in\Im\IH$,
furthermore $v^2=-1=w^2$ and thus
\begin{eqnarray*} \varphi_u(v)
&=& -\frac{-v-2w+(2\langle -v,w\rangle w+v)}{2+2\langle v,w\rangle}
\qquad=\quad w \enspace .
\end{eqnarray*}
The $\IR$-algebra homomorphism property ensures that
~$q\big(\varphi_u(x)\big)=\varphi_u\big(q(x)\big)$~ for
any polynomial $q$ with real coefficients and $x\in\IH$.
In particular, $\IR$-linearity yields
$$ y=\varphi_u(x) \quad=\quad
  \varphi_u\big( \Re(x)+|\Im(x)|\cdot v\big)
      \quad=\;\; \Re(x)+|\Im(x)| i \quad\in\quad \IR+i\IR $$
\qed 
\end{appendix}
\end{document}